\documentclass{edp-conf}
\usepackage{graphicx}
\begin{document}

\TitreGlobal{SF2A 2003}

%%-----------------------------
%%      the top matter
%%-----------------------------
\title{Radio Detection of Cosmic Ray Extensive Air Showers: present status of the CODALEMA experiment}
\author{A. Bell\'etoile$^1$}
\author{D. Ardouin$^1$}
\author{D. Charrier$^1$}
\author{R. Dallier}\address{SUBATECH, In2p3-CNRS/Universit\'e de Nantes/Ecole des Mines de Nantes}
\author{L. Denis}\address{Station de Radioastronomie de Nan\c{c}ay}
\author{P. Eschstruth}\address{LAL, In2p3-CNRS/Universit\'e de Paris Sud, Orsay}
\author{T. Gousset$^1$}
\author{F. Haddad$^1$}
\author{P. Lautridou$^1$}
\author{A. Lecacheux}\address{LESIA, Observatoire de Paris-Meudon}
\author{D. Monnier-Ragaigne$^3$}
\author{A. Rahmani$^1$}
\author{O. Ravel$^1$}
\runningtitle{Radiodetection of Cosmic Ray Air Showers}
\setcounter{page}{1}
% Keep this line, even if the page will be settled afterwards..

\index{Ardouin, D.}
\index{Bell\'etoile, A.}
\index{Charrier, D.}
\index{Dallier, R.}
\index{Denis, L.}
\index{Eschstruth, P.}
\index{Gousset, T.}
\index{Haddad, F.}
\index{Lautridou, P.}
\index{Lecacheux, A.}
\index{Monnier-Ragaigne, D.}
\index{Rahmani, A.}
\index{Ravel, O.}
% Repeat the authors here, this will help to make the final index

\maketitle
\begin{abstract}
Data acquisition and analysis for the CODALEMA experiment,
in operation for more than one year, has provided improved knowledge
of the characteristics of this new device.  At the same time, an
important effort has been made to develop processing techniques
for extracting transient signals from data containing interference.

% In operation for more than one year, data acquisition and analysis on the
% CODALEMA experiment has allowed to improve the knowledge with this new device.
% At the same time, an important effort has been made to develop signal processes
% able to extract transient signal from interferences.

\end{abstract}
%
%%-----------------------------
%%      your text
%%-----------------------------
\section{Motivation}
First experimented in the 60's (Linsley 1963, Jelley 1966, Weekes 2001),
radiodetection of extensive air showers (EAS) was set aside until recently when interest revived due to the evolution of electronic devices. Two experiments, namely CASA-MIA (Green et al. 2003) and KASKADE (Falcke et al. 2003), have attempted to detect radio-frequency transients associated with showers using antennas triggered by a particle detector array.  Up to now, no EAS candidates have been published. 

\noindent{}As the EAS develops in the atmosphere, several mechanisms of electromagnetic emission contribute to generate an electric pulse (Allan 1971). Considering only one of those mechanisms (Ravel et al. 2004), the signal resulting from a vertical ultra high energy cosmic ray shower with a primary particle energy of $10^{20}$~eV should reach 40 to 300~${\mu}$V$\times$m$^{-1}$ at a distance of 1~km, with a 10~ns to more than 1~${\mu}$s duration depending on the impact parameter. Lower energy showers would produce comparable signals at smaller distances. Such signals are readily measurable, and using pulse arrival times on at least 3 antennas, the incidence of the transient wavefront can be determined. In addition, it should be possible to extract information related to the primary particle (energy, impact parameter or nature) directly from the shape of the electric signal as a function of time.

\section{The CODALEMA setup}
Running by March 2003, the CODALEMA (COsmic ray Detection Array with Logarithmic ElectroMagnetic Antennas) experiment was set up at the Nan\c{c}ay Radio Observatory where suitable antennas (used in the 1-100~MHz frequency band for CODALEMA) were available in the DecAMetric array (DAM), an instrument dedicated to observing the sun and Jupiter. CODALEMA uses 6 of the 144 log-periodic antennas constituting the DAM as shown in the figure below.  

\begin{figure}[h]	\centering	\includegraphics[width=10.cm, height=3.cm]{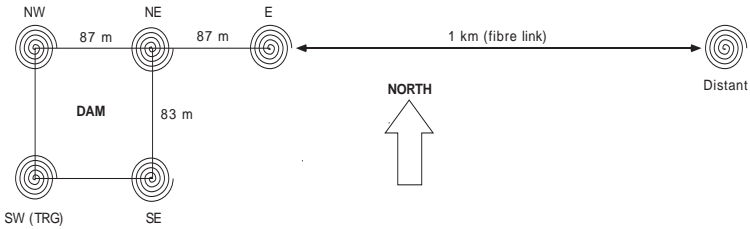}	\caption{ \it Current CODALEMA setup.The SW antenna is used to trigger.} 
	\label{codalema}\end{figure}

\noindent{}The 5 grouped antennas are linked, after signal amplification (gain 37 dB), via 150~m long cables to digital oscilloscopes (8 bit ADC, 500~MS/s, 10~${\mu}$s recording time). The data are stored on disk for off-line analysis. The $6^{th}$ antenna requires a quite different signal transmission technique (1~km optical fibre link), due to its location 1~km to the east. This particular antenna layout may make it possible to see the effect of the impact parameter on the signal (shape and amplitude). 
The originality of CODALEMA resides in the fact that the system is self-triggered using a devoted antenna filtered in an appropriate, interference free frequency band (33-65~MHz). Unlike the experiments mentioned above, a particle detector is not used to furnish a trigger. Signals from the whole set of antennas are acquired when a voltage threshold is reached on the trigger antenna (Dallier et al. 2003).

\section{Source localization}
Due to the presence of signals from radio transmitters particularly in the lower and upper parts of the full frequency band, applying an off-line numerical passband 33-65~MHz (trigger frequency band) is needed to bring a transient to light. Provided that a transient signal from a physical source is present in the data, this process can be used to observe coincidences involving several antennas. Assuming that the source of the signal is far enough from the detector, the propagating wavefront can be considered to be a plane. Thus, using the time delays of the signals from the different antennas, the location of the emission source can be reconstructed.

\begin{figure}[h]
	\centering
	\includegraphics[width=5.cm, height=5.3cm]{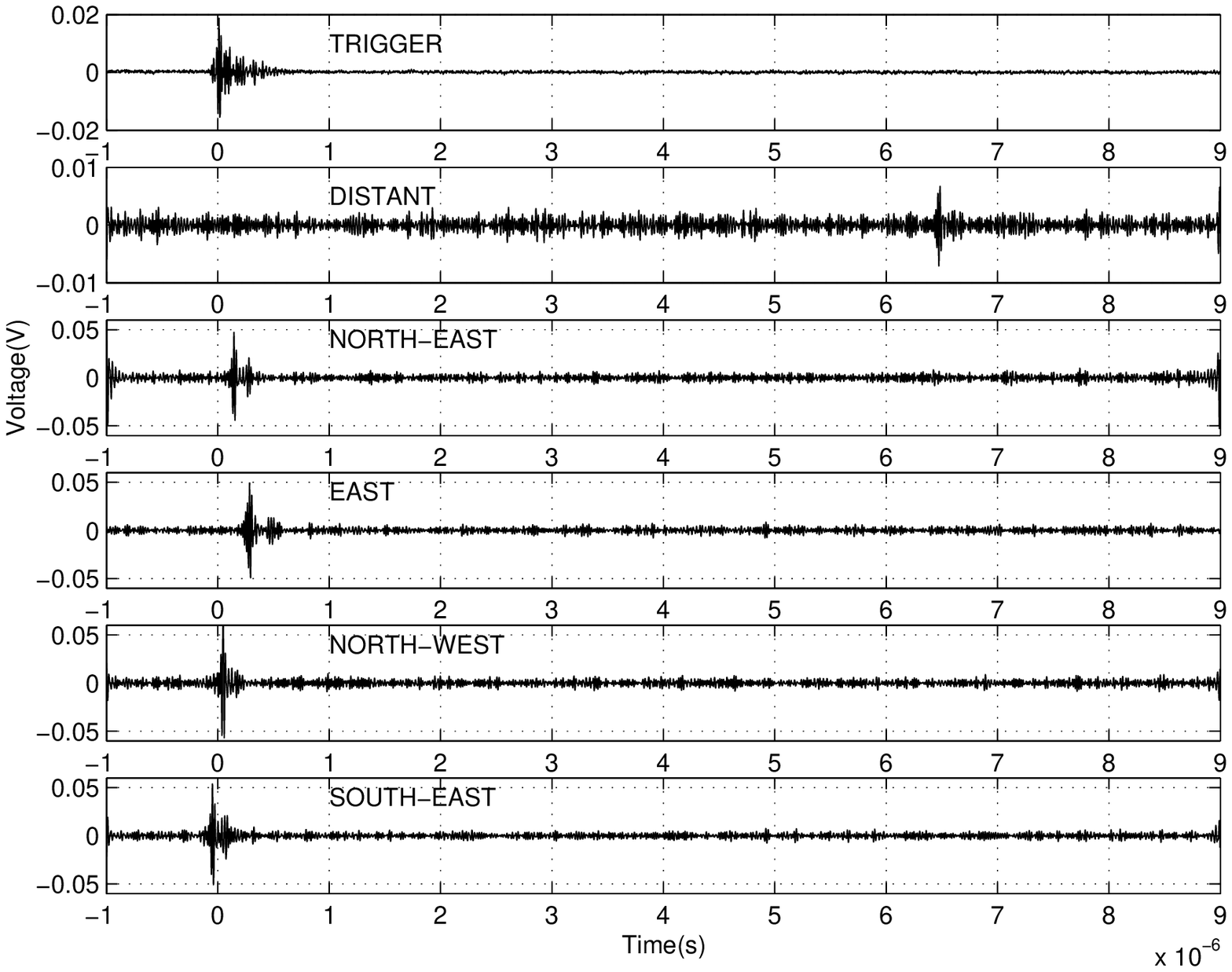}
	\includegraphics[width=5.cm, height=5.5cm]{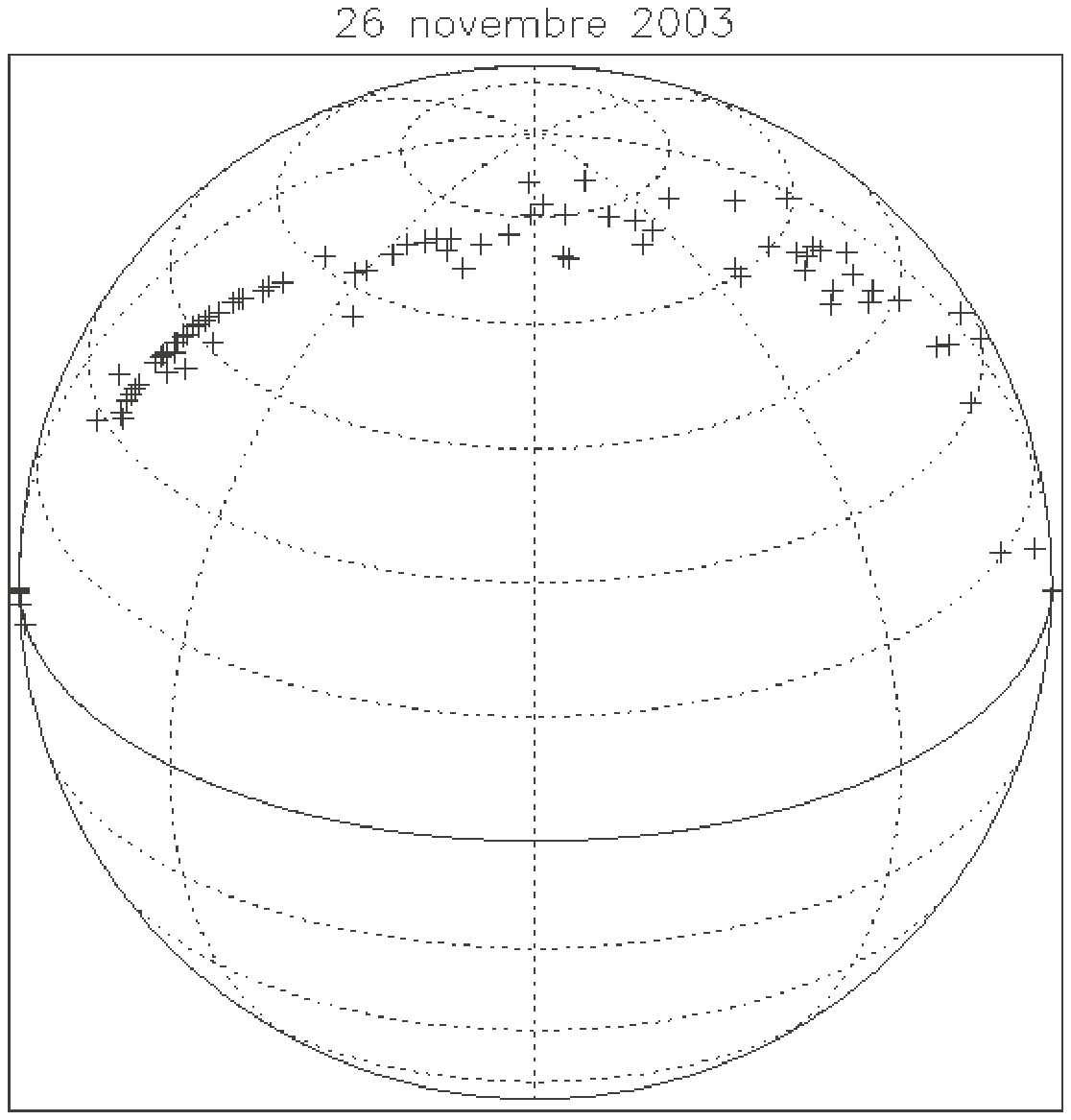}
	\caption{ \it Left : Coincidence transients detected with CODALEMA after 33-65~MHz numerical filtering. From Top to Bottom : trigger, distant (1~km), NE, E, NW and SW antennas respectively. Right : Location in the sky of sources of successive events.} 
	\label{results}
\end{figure}
	
\noindent{}An illustrative example of this is given in Fig.~\ref{results}, where a series of successive triggers occurred over a two minute period. For each event, a wavefront direction was calculated in terms of azimuth and elevation angles. Then they were plotted on a 3D graphic representing the sky over the Nan\c{c}ay Radio Observatory. From the duration of this series of triggers, the assumption could be made that they were from the same source moving from north to south above the detector array as an airplane could have done. Identifying such sources of anthropic interference makes it possible to reduce the counting rate. The rate obtained from the data after removing such contributions is compatible with expectations for EAS rates.

\section{Pulse extraction}
\begin{figure}[h]
	\centering
	\includegraphics[width=8.cm, height=4.cm]{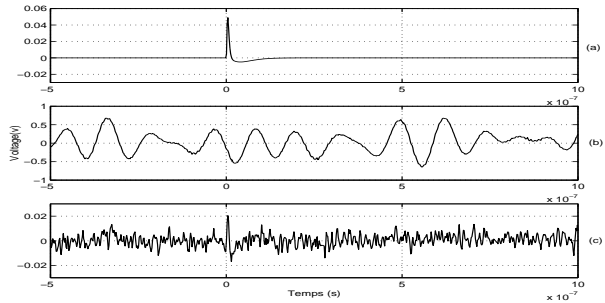}
	\caption{ \it Pulse extraction process applied to a simulated signal. (a) Original generated pulse (50~mV peak, i.e. about 200 ~${\mu}$V$\times$m$^{-1}$ electric field strength, 20~ns duration). (b) Simulated signal from adding the pulse to a background radio noise waveform recorded by CODALEMA. (c) Pulse recovered after signal processing (20~mV peak, 18~ns duration).} 
	\label{simulation}
\end{figure}

\noindent{}Although numerical filtering is useful for time tagging and source localization, it is not well adapted for providing information about the shape of the original electrical pulse. Therefore a signal processing tool able to extract the pulse waveform from the background radio noise has been developed. The method, based on the Discrete Fourier Transform technique (DFT), performs amplitude limiting on the strongest radio transmitter lines in the frequency domain (1-20 and 80-100~MHz). Then an appropriate spectrum model is fitted to the resulting spectrum in the 20-80~MHz frequency band. Finally, the whole-band fitted spectrum (1-100~MHz) associated with its phase is brought back into the time domain to get the pulse waveform. In order to evaluate the efficiency of such a process, simulated signals have been generated by adding a pulse to a typical background noise record from CODALEMA. An example of this process is shown Fig.~\ref{simulation} : the pulse (Fig.~\ref{simulation}a), hidden by radio interference (Fig.~\ref{simulation}b), is extracted (Fig.~\ref{simulation}c) without introducing oscillation in its structure. Notice, however, that the peak voltage is significantly reduced.

\noindent{}The assumptions that must be made when the spectrum is fitted put limits on the interest of the method. Another approach, based on Linear Predictive Coding filtering, is currently under investigation. An advantage of this second method is that no hypothesis is made concerning the detected transient. About 200000 recorded events, covering 18 months of activity, are currently under analysis.

\section{New developments}
The CODALEMA experiment will benefit from several upgrades in September 2004. In particular, seven more antennas will be added on a 600~m east-west line crossing the existing array. These are primarily intended for studying the influence of the impact parameter on EAS signals. In addition, the antenna array will be coupled to four double scintillators of 2.3~m$^2$ each, located with 100 m separation at the corners of the DAM array. Still triggered with the dedicated antenna, this combined setup should be capable of providing unambiguous evidence of EAS radio signals through the simultaneous detection of shower particles.

% \noindent{}The CODALEMA set up will know a couple of upgrades during September 2004.
% First, 7 extra antennas are going to be added to the existing array. They will be
% arranged on a east-west 600~m line just beside the original array. This should allow
% to study the influence of the impact parameter. Then, the antenna array is going to
% be coupled with 4 double scintillators of 2.25~m$^2$ each, arranged on a 100~m side
% square. Still triggered with the dedicated antenna, this new set up is expected to
% bring an unambiguous EAS radio signal.

%%-----------------------------
%%      your bibliography
%%-----------------------------


\begin{thebibliography}{99}
\bibitem{Volcano} J. Linsley, 1963, Phys. Rev. Lett. 10, 146.
\bibitem{Jelley} J.V. Jelley {\em et al.}, 1966, Nuovo Cimento A46, 649.
\bibitem{Weekes} T. Weekes, 2001, {\em in : Radio Detection of high Energy Particles}, ed D.Saltzberg  and  P.Gorham, Proceeding of first Int. Workshop RADHEP 2000, $N^o$ 579, AIP, Melville, NY, 3.
\bibitem{Allan} H.R. Allan, 1971, {\em Prog. in  Elem. part. and Cos. Ray Phys.}, ed J.G Wilson and S.A. Wouthuysen, (N. Holland Pub. Co.), Vol. 10, 171.
\bibitem{Green} K. Green {\em et al.}, 2003, Nucl. Inst. And Meth. A, 498.
\bibitem{Falcke} H. Falcke, P. Gorham, 2003, Astropart. Phys., Vol. 19, 477.
\bibitem{Ravel} O. Ravel {\em et al.}, 2004, Nuclear Instruments \& Methods in Physics Research - A 518, iss. 1-2, p213-215.
\bibitem{Dallier} R. Dallier {\em et al.}, 2003, {\em SF2A 2003 Scientific Highlights}, ed F. Combes, D. Barret, T. Contini and L. Pagani, published by EDP Sciences.

\end{thebibliography}
\end{document}